\documentclass[10pt,twocolumn,prd,aps,amssymb,amsmath]{revtex4}
\usepackage{graphics}
\usepackage[pdftex]{graphicx}

\newcommand{\ket}{\rangle}
\newcommand{\bra}{\langle}

\begin{document}
\preprint{}

\title{The Bayes cost in the binary decision problem}

\author{Bernhard K. Meister}
\email{bernhard.k.meister@gmail.com}
\affiliation{ Department of Physics, Renmin University of China, Beijing, China 100872}

\date{\today }

\begin{abstract}

The problem of quantum state discrimination between two wave functions  on a ring is considered. The optimal minimum-error probability 
is known to be given by the Helstrom bound. A new strategy is introduced by inserting either adiabatically or instantaneously an impenetrable barrier. The insertion point, independent of the shape of the initial wave function,  becomes a node. 
The resulting  modified wave functions  can be, if the initial functions are judiciously chosen,  distinguished with a smaller error probability, and as a consequence the Helstrom bound can be violated under  idealised conditions.   
\end{abstract}
\maketitle


\section{Introduction}
\label{sec:1a}

Experimental design and data analysis are common challenges in science, and particular acute in quantum mechanics. 
In the literature different approaches are discussed. 
In the paper  the Bayes procedure for state discrimination with the aim to minimise the expected cost is employed, since the existence of a prior associated with the states to be distinguished is assumed. 

 Two disparate concepts are combined in the paper: quantum state discrimination and the modification of the  quantum potential resulting in a transformation of the wave functions. 
 Bayesian hypothesis testing, the particular form of quantum state discrimination discussed here, was developed by Helstrom and others \cite{helstrom, holevo, yuen}. For a recent paper on the topic of state discrimination  between two possible states with given prior and transition probability see Brody {\it et al.} \cite{dbbm}. It is generally accepted, but will be challenged in the paper, that the optimal Bayes cost in the binary case, is given by the Helstrom bound, which only depends on the prior and the transition probability between the states and can be written in a simple closed form. 
The second concept is that wave functions can be modified in a beneficial way, if one inserts an impenetrable barrier corresponding to a potential spike in a simple configuration space, here chosen to be either one or two rings.
   In an earlier paper \cite{bkm2011} these two ideas were applied to a one-dimensional infinite square well. The current approach is simpler due to the rotational symmetry of the ring, which is only broken once the barrier is inserted, and the eigenvalue degeneracy of the Hamiltonian. 


Next, a description of the setup and the procedure to be analysed. We distinguish  the instantaneous and the adiabatic case.  These are distinguished by the configuration space used, either the  wave function is spread over one or two rings, and by the barrier insertion speed.
What both cases have in common is that we are presented with the following decision problem. With equal probability, prior of $1/2$,
one of two quantum states is put into the configuration space. Our challenge is to find out, which state has been selected. 
Two strategies for calculating the Bayes cost are proposed. In the first strategy,  the combination of prior and transition probability between the two quantum states alone is sufficient to calculate the conventional optimal minimum error probability, i.e. the Helstrom bound, prior to the insertion of any barrier. 
The standard procedure, reliant on the optimal POVM as described in the book by Helstrom\cite{helstrom}, results in the following binary decision   cost 
\begin{eqnarray}
\frac{1}{2}-\frac{1}{2}\sqrt{1-\cos^2(\delta)},\nonumber
\end{eqnarray}
where $\cos^2(\delta)$ is the transition probability between the two candidate states, and cost $1$ is assigned to an incorrect  and cost $0$ for a correct decision.

  The second  strategy for calculating the error probability is novel. The wave function is first modified by barrier insertion breaking the symmetry of the configuration space (either one or two rings).  
  The insertion of the barrier can be carried out with different speeds. We consider the two extreme cases: instantaneous and adiabatic. This insertion can require energy and can modify the wave function, since the amplitude of the wave function at the impenetrable barrier location has to be zero. The modified wave function is then probed and the  new binary decision cost estimated, where either  the added information gained about the energy required for insertion (in the instantaneous case) is taken into account or the change in the transition probability of the modified candidate states (in the adiabatic case) is noticed. In both situations the Helstrom bound can be broken.






There are two main motivations for the work presented here. On the one hand it will possibly shed some light on foundational issues in quantum measurement theory, and on the other hand  a plethora of problems in quantum information theory depend on optimal state discrimination.

The bare essentials about quantum mechanics on the ring are given next. 
 A particle of mass $m$ trapped on a ring of radius one has the Hamiltonian 
\begin{eqnarray}
H= - \frac{\hbar^2}{2m }\frac{d^2}{d\theta^2}.\nonumber
\end{eqnarray}
 with energy eigenvalues $E_n=\frac{ \hbar^2 n^2}{2 m}$ for $n\in \mathbb{N}$ .
The wave function is defined for $\theta\in[0,2 \pi] $.

 


The structure of the rest of paper is as follows. In section II and III  the impact of an insertion of a barrier on a ring is studied in the instantaneous and adiabatic case respectively.  The  binary choice problem  between two quantum states is  tackled at the end of each section. 
In the conclusion the result is briefly reviewed  and some general comments added.





 \section{Instantaneous insertion of a barrier }
\label{sec:1b}

In this section the barrier insertion at both nodal and non-nodal points is considered to be instantaneous. 
The nodal point insertion is dealt with first. This is easier,
since the particle wave function and energy is left unchanged -  for background see section II of  Bender {\it et al.} \cite{bbm}, where 
 a series of results for a particle in a one dimensional box are established, which are directly applicable to quantum mechanics on a ring.
As an aside, we only call a point a node, or more correctly a `fixed node', if the amplitudes  at this point stays zero at all times. 
Wave functions that are superposition of eigenfunctions of $H$ can have zero amplitude points that change with time. These we do not consider, since an insertion at a `transitory node' can require energy.

 

The situation is  more intricate  for an insertion at a non-nodal point.
Energy is needed to modify the wave function, if the insertion happens at a point, where the amplitude is non-vanishing. In the idealised setting considered here the required energy  is infinite. The energy localised in
the barrier point inserted at $t=0$ propagates through the system at $t > 0$
and increases the energy on the ring. The result
is a fractal wave function. The details of the calculation can be found in section  IV \& VI of Bender {\it et al.} \cite{bbm} and and section II of \cite{bkm2011}.




In the following paragraphs the cost is evaluated before and after  the insertion of a barrier for distinguishing the following two candidate wave functions 
\begin{eqnarray}
\phi(\theta)&:= &\frac{1}{\sqrt{\pi}}\sin(  \theta ),\nonumber\\
\psi(\theta)&:=&  \frac{1}{\sqrt{\pi}}\sin(\theta + \alpha),\nonumber
\end{eqnarray}  
where $\theta \in [0, 2 \pi]$, and $\alpha\in(0,\pi/2)$. 
The initial overlap of the wave functions is
 \begin{eqnarray}
| \bra \phi | \psi \ket |^2= 
\frac{1}{\pi^2} \Big|\int_{0}^{2\pi} d \theta \sin ( \theta )\sin( \theta + \alpha)\Big|^2 =\cos^2(\alpha).\nonumber
\end{eqnarray}
Due to the pre-insertion symmetry of the set-up both candidate wave functions are eigenfunctions of the Hamiltonian. The symmetry is only broken by  the barrier. The barrier is inserted at the point $\pi$ at the time $t=0$ and leaves $\phi$ unchanged. 
The situation is different for $\psi$, since it lacks a node at this point.

The new wave functions, still taken at $t=0$ to avoid unnecessary exponential factors that do not affect the transition probability, are therefore
\begin{eqnarray}
\phi'(\theta)&:= &\frac{1}{\sqrt{\pi}}\sin(  \theta ),\nonumber\\
\psi'(\theta)&:=&  \frac{1}{\sqrt{\pi}}\sin(\theta + \alpha),\nonumber
\end{eqnarray}
with $\theta \in [0, 2 \pi]$, but excludes the isolated point $\pi$, where both wave functions are zero.
The transition probability stays unchanged \begin{eqnarray}
| \bra \phi' | \psi' \ket |^2
=\frac{1}{\pi^2} \Big|\int_{0}^{\pi} d \theta \sin ( \theta )\sin( \theta + \alpha)\nonumber\\
+\int_{\pi}^{2\pi} d \theta \sin ( \theta )\sin( \theta + \alpha)\Big|^2 =\cos^2(\alpha),\nonumber
\end{eqnarray}
since the new amplitudes, except for a set of measure zero,
match the original values.

Does this suggest an inability to break the Helstrom bound, since the transition probability remains unchanged? This is not the case, because there is an additional source of information about the states given by the energy required to insert the barrier. 

Two potentially controversial assumption underpin this process.
 The first assumption is tied to the insertion of an impenetrable barrier of infinitesimal width.  
It is an idealisation, but still has some experimental relevance, since 
 a slower, but not infinitely slow, insertion of a barrier also requires a finite amount of work. The finite speed case is harder to analyse and  demands the solution of a time-dependent Hamiltonian. 
The second assumption is that one is able to measure the energy needed to insert a barrier.  A fuller analysis requires a realistic interpretation of what    it means to insert a barrier, e.g. change the potential with time or let a laser beam interact with the test particle, and how the barrier interacts with the wave function. 

 After the insertion there are two sources of accessible information. The first is associated with the
transformed wave function. The new wave function can be tested using conventional Helstrom strategies to reconfirm the standard Helstrom bound.
There is a novel source of information obtained from the measurement of the energy needed to insert the barrier. This second and new type of information provides supplementary insight, since it differs markedly for the two candidate wave functions.
If the first test function $\phi$ is on the ring, the needed insertion energy is zero. If the second test function $\psi$ is on the ring, a substantial amount of energy, infinite in the idealised instantaneous case, is necessary.
This additional `side information' will always be 	advantageous  and help to  decrease the cost below the Helstrom bound. 

\section{Adiabatic insertion of an impenetrable barrier}

After the instantaneous insertion discussed above, we turn to the other extreme:  adiabatic insertion.
Like the previous case it represents an idealisation, since the system is assumed to never depart from equilibrium. This requires the process to be infinitely slow  
and the change in the energy of the system to be minimised. As before, if the barrier point is not a node, 
 the state of the wave function is liable to be modified.
For background about an adiabatic insertion see  section VII of  Bender {\it et al.} \cite{bbm}. As before, the results for an infinite potential well 
are directly transferable to quantum mechanics on a ring.


The configuration space of the last section is extended to an assembly of two rings.
The doubling of the configuration spaces  simplifies the analysis, since we  can select a  wave function, where the amplitude on the second ring is the negative of the amplitude on the first ring after the barriers are inserted.
The candidate wave functions
are now chosen to be of the form 
  \begin{displaymath}
  \widehat{\phi} :=\left\{ 
    \begin{array}{cc}
    \frac{1}{\sqrt{2\pi}}  \sin( x ) & 0\leq x \leq 2 \pi \\
     \frac{1}{\sqrt{2\pi}} \sin(  x) & 2 \pi < x<4\pi \\
         \end{array}
         \right.
  \end{displaymath}
and
 \begin{eqnarray}
 \widehat{\psi}: =\left\{ 
    \begin{array}{cc}
   \frac{1}{\sqrt{2\pi}}    \sin( x +\alpha) & 0\leq x \leq 2 \pi \\
      \frac{-1}{\sqrt{2\pi}}\sin( x+\beta) & 2 \pi <x< 4\pi, \\
         \end{array}\nonumber
         \right.
\end{eqnarray}
where  the interval $[0,2 \pi]$ corresponds to the first ring, $[2 \pi, 4\pi]$  to the second ring, and $\alpha, \beta \in (-\pi/4,\pi/4)$. The pre-insertion transition probability is 
 \begin{eqnarray}
| \bra \widehat{\phi}| \widehat{\psi} \ket |^2=\frac{1}{4}\Big(\cos(\alpha)- \cos(\beta)\Big)^2.\nonumber
\end{eqnarray}

The adiabatic insertion is now carried out simultaneously on both rings at $\pi$ and $3\pi$ at $t=0$. As a result, $\widehat{\phi} $ is left unchanged, but $\widehat{\psi}$ is modified and turns into
 \begin{eqnarray}
 \widehat{\psi}' =\left\{ 
    \begin{array}{cc}
   \frac{1}{\sqrt{2\pi}}    \sin(x  ) & 0\leq x \leq 2 \pi\\
      \frac{-1}{\sqrt{2 \pi}}\sin(  x ) & 2 \pi < x< 4 \pi.\\
         \end{array}\nonumber
         \right.
\end{eqnarray}
The fact that $\widehat{\phi}$ is left unchanged is not surprising, since both insertion points are nodes. 

The change of $\widehat{\psi}$ needs to be studied carefully, and depends on how one defines a `minimal energy transformation'. In our case, a minimal energy change, requiring no energy at all, is achieved by a rotation of the state. Next the question arises, why the new state is given  positive and negative amplitude at various points. Our choice maximises the overlap between the state before and after the transformation, while not violating the two constraints. First, the insertion points become nodes. Second, the transformation requires minimal energy. These choices seem to be  a reasonable way to determine the transformation, but experiments will be the final arbiter.

Prior to insertion, the standard optimal lower bound in the binary decision problem is given by the Helstrom bound
\begin{eqnarray}
\frac{1}{2}-\frac{1}{2}\sqrt{1-\frac{1}{4}\Big(\cos(\alpha)-\cos(\beta)\Big)^2}.\nonumber
\end{eqnarray}
The cost after the insertion is even  simpler to calculate,  
since the candidate wave functions after modification are orthogonal, and therefore the new binary cost is zero. 
In the upcoming conclusion the results are reviewed and some related issues noted.

%

\section{Conclusion}

The aim of the paper was to show that the Helstrom bound in the binary quantum discrimination setting can be breached.
Inserting a barrier instantaneously in a ring at a non-nodal point always requires energy. This provides auxiliary information and can be used to help distinguish between different states. This result can be extended beyond instantaneous insertion, since any non-adiabatic insertion at a non-nodal point, but not at a fixed nodal point, needs energy. 
In the adiabatic case the minimal energy transformation associated with the insertion at a non-nodal point leads to a shift of the wave function changing the transition probability between candidate states. In both cases the binary  cost can be lowered below the Helstrom bound.

The exact cost reduction in the non-adiabatic case depends on how precisely one can place the barrier, the width of the barrier, the speed of insertion, i.e. three issues related to time evolution of the potential associated with the barrier, and the measurement uncertainty associated with the energy needed to insert the barrier.
If each of these four points can be addressed satisfactorily, then not only in the idealised situation can one obtain an improvement of the optimal cost beyond the  Helstrom bound.

Naturally, one can criticise the failure to provide a realistic time evolution of the barrier as well as the lack of an explicit calculation of the time evolution of the  state. Only the infinitely slow and fast insertion were evaluated. In defence, one can point to the idealised nature of the proposal and the statement that more realistic examples are an interpolation between the  adiabatic and instantaneous extreme. 

The following Gedankenexperiment might be instructive, since it shows that it is possible to  construct an example were no information is transferred to the experimenter, when the potential representing the barrier is changed. 
Imagine an experimenter doing a fixed amount of work per unit of time to insert the barrier, i.e. pushes in the barrier with constant power.  
   Dependent on the test wave function the change of the potential is either larger or smaller. The potential takes on different shapes. 
The changing potentials affect the states in different ways, ergo can change the distance between the states, without direct, energy based, leakage of information to the outside.
 
 A critical point, which hampered the analysis of the particle in a box in an earlier paper \cite{bkm2011}, was the reliance on superpositions of eigenfunctions. This led to energy-level dependent entanglement of the candidate wave function    with the barrier and influenced the transition proability of the states. Each element of a superposition of energy eigenstates gained during the barrier insertion  a different amount of energy. 
This can be avoided in the new configuration space with added symmetry and the beneficial degeneracy of the energy eigenvalues. 

 As an aside,  the instantaneous case can be extended trivially to the two ring configuration, and any possible two ring configuration can be mapped into a one ring configuration.



One could ask, if this result suggests the possibility of superluminal communication. The adiabatic nature of the insertion in the second case prevents the process from being exploited for communication. In general, the Schr\"{o}dinger equation, a diffusion equation without a propagation speed limit, already indicates that one should not expect too much from non-relativistic quantum mechanics.

Almost without fail quantum algorithms can be viewed as procedures to distinguish between different states.  The method described above works for states with arbitrary overlap and should find application in the area of quantum algorithm. This is  to be examined in a separate paper. 

 Experimental implementation, for example in the area of Bose-Einstein condensate or ion traps,   with a laser beam as a barrier, should be of interest. 

\hspace{-.38cm}The author wishes to express his gratitude to D.C. Brody for  stimulating discussions.

%

\end{document}